\documentclass{article}

\usepackage[preprint]{neurips_2024}

\usepackage[utf8]{inputenc}
\usepackage[T1]{fontenc}
\usepackage{hyperref}
\usepackage{url}
\usepackage{booktabs}
\usepackage{amsfonts}
\usepackage{nicefrac}
\usepackage{microtype}
\usepackage{xcolor}
\usepackage{graphicx}
\usepackage{amsmath}
\usepackage{amssymb}
\usepackage{float}
\usepackage{longtable}
\usepackage{multirow}
\usepackage{adjustbox}
\usepackage{natbib}

\title{scBench: Evaluating AI Agents on Single-Cell RNA-seq Analysis}

\author{
  Kenny Workman \hspace{1.5em}
  Zhen Yang \hspace{1.5em}
  Harihara Muralidharan \hspace{1.5em}
  Aidan Abdulali \hspace{1.5em}
  Hannah Le \\[1em]
  LatchBio, San Francisco, CA \\[0.5em]
  Correspondence: \texttt{kenny@latch.bio}
}

\begin{document}
\raggedbottom

\maketitle

\begin{abstract}
As single-cell RNA sequencing datasets grow in adoption, scale, and complexity, data analysis remains a bottleneck for many research groups. Although frontier AI agents have improved dramatically at software engineering and general data analysis, it remains unclear whether they can extract biological insight from messy, real-world single-cell datasets. We introduce scBench, a benchmark of 394 verifiable problems derived from practical scRNA-seq workflows spanning six sequencing platforms and seven task categories. Each problem provides a snapshot of experimental data immediately prior to an analysis step and a deterministic grader that evaluates recovery of a key biological result. Benchmark data on eight frontier models shows that accuracy ranges from 29--53\%, with strong model-task and model-platform interactions. Platform choice affects accuracy as much as model choice, with 40+ percentage point drops on less-documented technologies. scBench complements SpatialBench to cover the two dominant single-cell modalities, serving both as a measurement tool and a diagnostic lens for developing agents that can analyze real scRNA-seq datasets faithfully and reproducibly.
\end{abstract}

\section{Introduction}

Single-cell RNA sequencing (scRNA-seq) is a workhorse assay in research
biology, providing transcriptional measurements at single-cell resolution to
interrogate molecular state of tissues. As datasets grow in size and
experimental usage broadens, drawing scientific conclusions increasingly
depends on multi-step and resource-intensive computational methods that bridge
techniques in statistics, high-dimensional data analysis, and programming. For
many research groups, analysis—not sequencing—has become a rate-limiting step~\citep{lahnemann2020challenges}.

Agents—large language models (LLMs) that write code, invoke tools, and iterate
toward a goal—have emerged with rapidly growing capabilities in software
engineering and data analysis~\citep{yang2024sweagent}. However, agents for scRNA-seq remain both
unreliable and underpowered, prone to scientific inaccuracies and
hallucinations, and frequently fail to complete domain-specific analysis steps
that depend on messy, real-world datasets.

Existing biology benchmarks emphasize recall, interpretation, or
literature-style reasoning~\citep{jin-etal-2019-pubmedqa,tinn2023finetune}, and do not require empirical interaction with data
or faithfully represent real-world analysis tasks. As a result, we lack a
standard, deterministic yardstick for data-grounded scRNA-seq analysis.

We introduce \textbf{scBench}, a benchmark of 394 verifiable problems distilled
from routine scRNA-seq workflows spanning six sequencing platforms and seven
task categories. Each evaluation consists of a data snapshot, a
natural-language task, and a deterministic grader. Across eight frontier models
evaluated under a common harness, the best model reaches 52.8\% accuracy, with
large task- and platform-dependent performance swings. Together with
SpatialBench for spatial transcriptomics, scBench provides a complementary
diagnostic for measuring and improving agent competence on the two dominant
transcriptional assays.

\section{Results}

\begin{table}[t]
\caption{Number of evaluations by platform and task category.}
\label{tab:inventory}
\centering
\small
\begin{tabular}{lrrrrrrrr}
\toprule
 & QC & Norm. & Dim.\ Red. & Clust. & Cell Typ. & Diff.\ Exp. & Traj. & Total \\
\midrule
BD Rhapsody  &  6 & 11 & 14 &  7 & 13 & 10 & --- &  61 \\
Chromium     & 10 & 11 & 15 &  8 &  5 & 11 & --- &  60 \\
CSGenetics   &  4 &  5 &  7 &  5 & 20 &  1 & --- &  42 \\
Illumina     &  8 &  7 & 10 & 12 & 33 &  8 &  7 &  85 \\
MissionBio   &  8 &  3 &  5 & 12 & 34 & 19 & --- &  81 \\
ParseBio     & --- &  7 & 18 &  5 & 13 & 22 & --- &  65 \\
\midrule
Total        & 36 & 44 & 69 & 49 & 118 & 71 &  7 & 394 \\
\bottomrule
\end{tabular}
\end{table}

\begin{figure}[t]
\centering
\includegraphics[width=\textwidth]{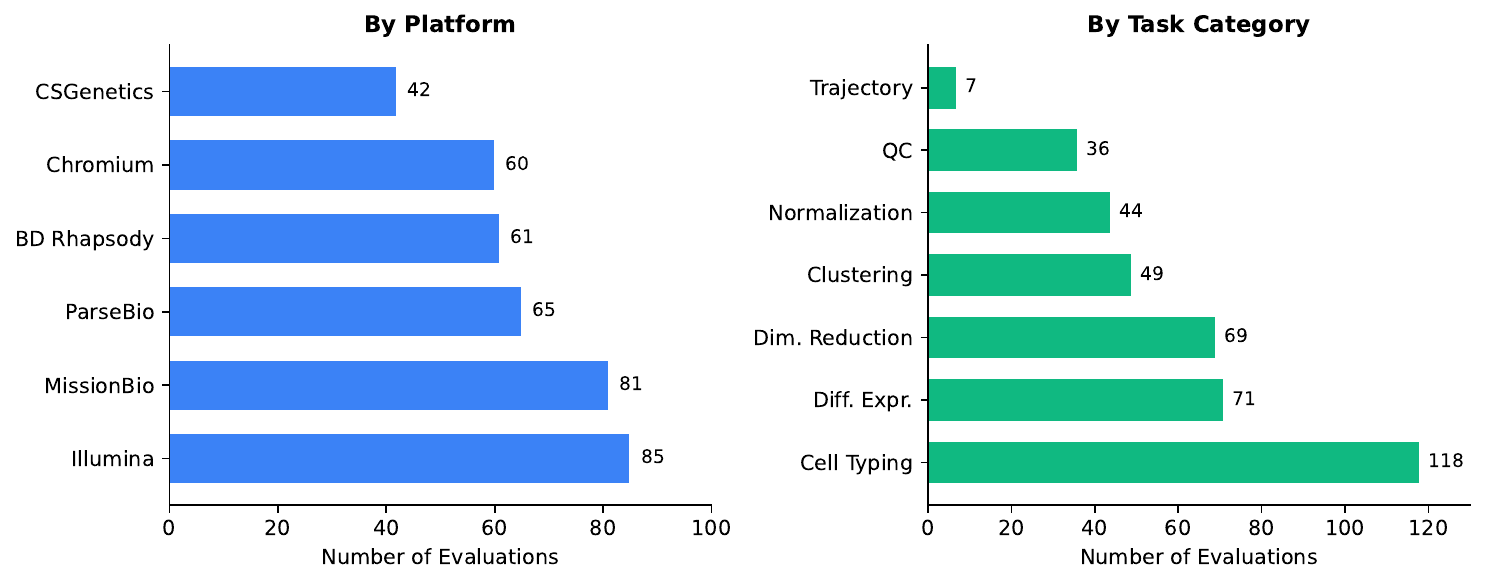}
\caption{Distribution of 394 evaluations across platforms and task categories. Cell typing and differential expression dominate; ParseBio lacks QC evaluations.}
\label{fig:inventory}
\end{figure}

\subsection{scBench: Verifiable Problems from Real Workflows}

scBench comprises 394 evaluations spanning six sequencing platforms and seven
task categories (Table~\ref{tab:inventory}). Each evaluation pairs a data
snapshot (often an AnnData \texttt{.h5ad} file) with a natural-language task
prompt and a deterministic grader that scores the agent's structured JSON
output as a pass or fail. The benchmark focuses on the analysis
stages with the greatest dataset-specific variation: cell typing (118
evaluations, 30\%) and differential expression (71, 18\%) together account for
nearly half the benchmark. Normalization (44) and QC (36) are smaller because
these procedural steps admit fewer distinct problem formulations per dataset.
Platform representation ranges from Illumina (85 evaluations) and MissionBio
(81) to CSGenetics (42). ParseBio lacks QC evaluations because its vendor
workflow omits explicit quality filtering, limiting cross-platform QC
comparisons to five platforms. MissionBio Tapestri is a targeted DNA+protein
platform rather than RNA-seq; we include it to stress-test whether agents
generalize beyond transcriptomic workflows to related single-cell analysis
patterns (clustering, cell typing from protein markers, variant interpretation).

The two axes, platform and task, allow stratified analysis of model
performance. Task categories reveal a gradient of accuracy: normalization
applies standard transformations often with well-understood implementations; an
agent need only identify the correct function call. Cell typing and
differential expression require multi-step reasoning and contextual scientific
judgement: selecting marker genes, interpreting cluster identity, subsetting
cells, choosing statistical tests, and identifying tissue-specific signatures.
Platform diversity tests generalization beyond training-data familiarity.
Chromium and Illumina dominate public repositories and tool documentation;
MissionBio and ParseBio appear less frequently, use non-standard data
structures, and sport lesser-known technical footguns. Models that overfit on
Scanpy tutorials without learning transferable analysis techniques should
collapse on underrepresented platforms.  Sections~\ref{sec:task}
and~\ref{sec:platform} quantify these effects.

\subsection{Aggregate Model Performance}

\begin{table}[t]
\caption{Overall model performance on scBench (394 evaluations, 3 replicates, mini-SWE-agent harness).}
\label{tab:models}
\centering
\small
\begin{tabular}{llrcc}
\toprule
Model & Provider & Accuracy (\%) & 95\% CI & Latency (s) \\
\midrule
Claude Opus 4.6   & Anthropic & 52.8 & (48.3, 57.2) & 303 \\
Claude Opus 4.5   & Anthropic & 49.9 & (45.3, 54.4) & 154 \\
GPT-5.2           & OpenAI    & 45.2 & (40.9, 49.5) & 133 \\
Claude Sonnet 4.5 & Anthropic & 44.2 & (39.9, 48.6) & 193 \\
GPT-5.1           & OpenAI    & 37.9 & (33.7, 42.0) &  94 \\
Grok-4.1          & xAI       & 35.6 & (31.6, 39.7) & 180 \\
Grok-4            & xAI       & 33.9 & (30.1, 37.8) & 203 \\
Gemini 2.5 Pro    & Google    & 29.2 & (25.6, 32.9) & 300 \\
\bottomrule
\end{tabular}
\end{table}

\begin{figure}[t]
\centering
\includegraphics[width=\textwidth]{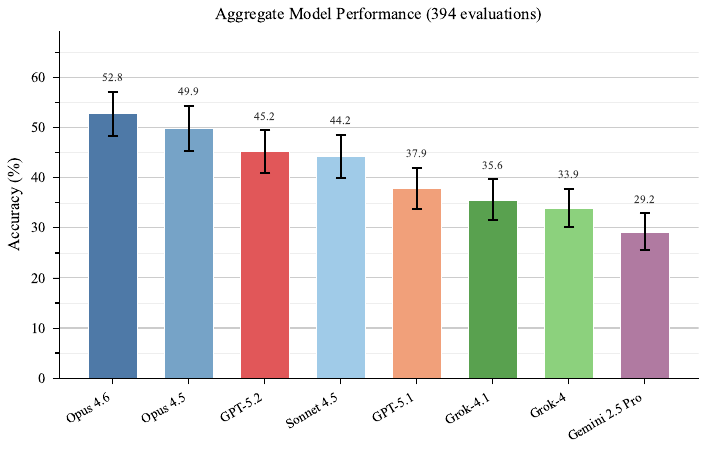}
\caption{Aggregate accuracy of 8 frontier models on scBench (394 evaluations, 3 replicates each). Error bars show 95\% confidence intervals computed via two-stage aggregation with the $t$-distribution.}
\label{fig:aggregate}
\end{figure}

\begin{figure}[t]
\centering
\includegraphics[width=\textwidth]{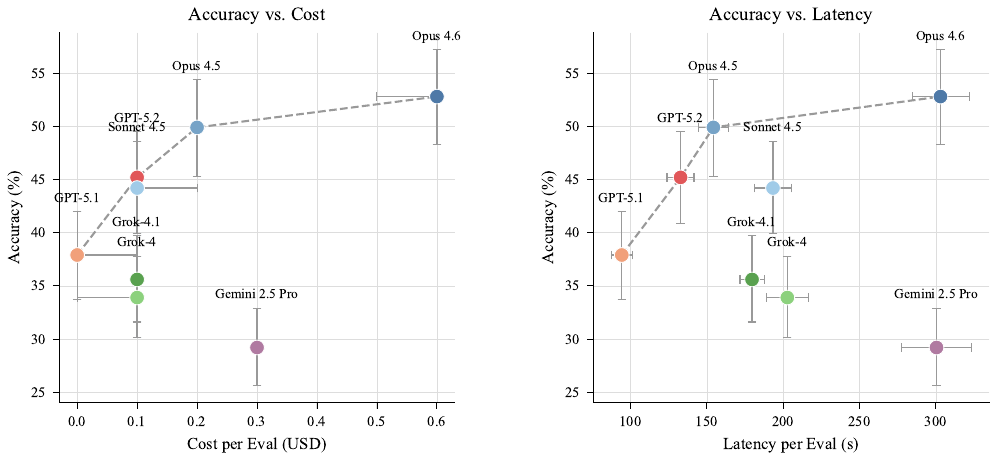}
\caption{Accuracy versus cost (left) and latency (right). Dashed lines connect Pareto-optimal models. GPT-5.2 achieves near-top accuracy at lower cost; Opus 4.6 leads accuracy but incurs higher cost and latency.}
\label{fig:pareto}
\end{figure}

We evaluated eight frontier models from four providers
(Table~\ref{tab:models}). Claude Opus 4.6 achieves the highest accuracy at
52.8\% (95\% CI: 48.3--57.2\%), followed by Claude Opus 4.5 at 49.9\% and
GPT-5.2 at 45.2\%. Claude Sonnet 4.5 reaches 44.2\%, placing fourth despite
being a smaller model. The bottom tier comprises GPT-5.1 (37.9\%), Grok-4.1
(35.6\%), Grok-4 (33.9\%), and Gemini 2.5 Pro (29.2\%).

The 23.6 percentage point spread between best and worst models exceeds
SpatialBench's 18.3 pp spread, indicating that scBench discriminates model
capability despite the higher overall accuracy. Anthropic models occupy the top
four positions, with both Opus variants and Sonnet outperforming all
competitors. Stratified analysis (Sections 2.3--2.4) reveals where models
diverge.

\subsection{Task Category Analysis}
\label{sec:task}

\begin{table}[t]
\caption{Accuracy (\%) by task category with 95\% CI. Best result per task in \textbf{bold}.}
\label{tab:bytask}
\centering
\footnotesize
\begin{adjustbox}{max width=\textwidth}
\begin{tabular}{lcccccc}
\toprule
Model & QC & Norm. & Dim.\ Red. & Clust. & Cell Typ. & Diff.\ Expr. \\
\midrule
Opus 4.6   & 61.1 (45.3, 76.9) & 82.4 (71.8, 93.0) & \textbf{55.4} (43.7, 67.1) & \textbf{52.7} (40.6, 64.9) & \textbf{48.2} (40.1, 56.2) & \textbf{41.4} (31.7, 51.0) \\
Opus 4.5   & \textbf{63.9} (48.1, 79.7) & \textbf{83.8} (73.1, 94.5) & 54.8 (42.8, 66.7) & 42.6 (29.5, 55.6) & 45.8 (37.6, 53.9) & 33.3 (23.8, 42.9) \\
GPT-5.2    & 63.0 (47.5, 78.4) & 74.8 (62.1, 87.4) & 50.0 (38.7, 61.3) & 42.6 (30.0, 55.1) & 39.5 (32.3, 46.8) & 29.9 (21.0, 38.8) \\
Sonnet 4.5 & 61.1 (45.7, 76.5) & 82.9 (71.6, 94.2) & 50.7 (39.8, 61.7) & 39.0 (26.9, 51.1) & 35.6 (28.2, 43.0) & 27.4 (18.5, 36.2) \\
GPT-5.1    & 60.2 (44.7, 75.6) & 62.2 (48.3, 76.1) & 47.3 (35.4, 59.2) & 33.3 (21.3, 45.4) & 29.1 (22.8, 35.4) & 25.2 (17.1, 33.4) \\
Grok-4.1   & 49.1 (34.2, 64.0) & 65.8 (52.3, 79.2) & 41.1 (30.0, 52.1) & 31.2 (20.5, 41.9) & 30.2 (23.6, 36.8) & 20.1 (12.3, 27.9) \\
Grok-4     & 40.7 (27.5, 54.0) & 51.4 (39.5, 63.2) & 38.7 (27.3, 50.1) & 34.8 (24.6, 44.9) & 29.1 (22.6, 35.6) & 25.2 (17.0, 33.4) \\
Gemini     & 47.2 (34.2, 60.3) & 59.5 (45.3, 73.6) & 35.1 (24.8, 45.4) & 29.8 (20.4, 39.2) & 22.0 (16.3, 27.8) & 13.7 (7.9, 19.4) \\
\bottomrule
\end{tabular}
\end{adjustbox}
\end{table}

\begin{figure}[t]
\centering
\includegraphics[width=\textwidth]{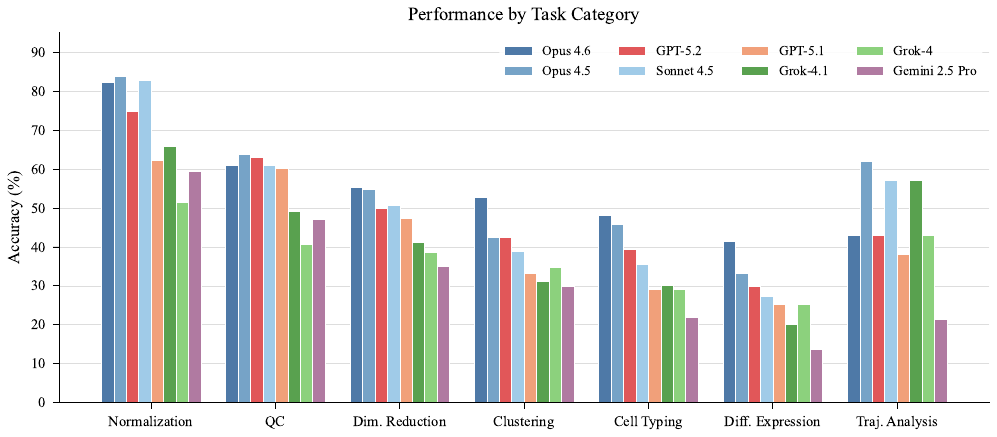}
\caption{Accuracy (\%) by model and task category. Tasks ordered by difficulty (normalization easiest, differential expression hardest). Error bars show 95\% confidence intervals. The difficulty gradient is consistent across models.}
\label{fig:task_comparison}
\end{figure}

Task categories reveal a consistent difficulty gradient (Table~\ref{tab:bytask},
Figure~\ref{fig:task_comparison}). Normalization is easiest (cross-model mean
70.4\%), followed by QC (55.3\%). These procedural tasks vary across biological
contexts but often involve applying
well-understood transformations. Differential
expression is hardest (mean 27.0\%), with cell typing (34.9\%) and clustering
(38.3\%) in the middle. Seven of eight models follow the same difficulty
ordering.

Differential expression is also most discriminative, with a 27.7 pp spread
between best and worst models. Model differences concentrate in judgment-heavy
stages---DE and cell typing---rather than procedural ones.

\subsection{Platform-Dependent Performance}
\label{sec:platform}

\begin{table}[t]
\caption{Accuracy (\%) by sequencing platform with 95\% CI. Best result per platform in \textbf{bold}.}
\label{tab:byplatform}
\centering
\footnotesize
\begin{adjustbox}{max width=\textwidth}
\begin{tabular}{lcccccc}
\toprule
Model & CSGenetics & BD Rhapsody & Illumina & Chromium & ParseBio & MissionBio \\
\midrule
Opus 4.6   & 74.6 (62.5, 86.7) & 53.0 (41.8, 64.2) & 52.5 (42.8, 62.3) & \textbf{51.7} (40.4, 62.9) & \textbf{53.2} (41.3, 65.1) & \textbf{42.0} (32.4, 51.6) \\
Opus 4.5   & \textbf{77.0} (65.9, 88.1) & \textbf{55.7} (44.7, 66.7) & 50.6 (40.4, 60.7) & 47.1 (35.5, 58.7) & 41.7 (28.2, 55.2) & 37.9 (28.3, 47.4) \\
GPT-5.2    & 65.1 (53.2, 77.0) & 54.6 (43.5, 65.8) & \textbf{54.5} (45.9, 63.1) & 46.6 (35.2, 57.9) & 35.9 (23.3, 48.5) & 23.0 (15.2, 30.9) \\
Sonnet 4.5 & 70.6 (57.8, 83.5) & 53.6 (42.1, 65.0) & 41.2 (31.9, 50.4) & 46.0 (35.4, 56.6) & 33.3 (22.3, 44.4) & 34.2 (25.0, 43.3) \\
GPT-5.1    & 46.8 (33.6, 60.0) & 41.5 (30.9, 52.2) & 50.0 (41.6, 58.4) & 46.0 (34.3, 57.7) & 25.0 (13.5, 36.5) & 20.2 (13.0, 27.3) \\
Grok-4.1   & 46.0 (33.5, 58.6) & 42.6 (31.8, 53.4) & 43.1 (34.1, 52.2) & 42.5 (32.0, 53.0) & 25.0 (13.4, 36.6) & 18.9 (13.0, 24.9) \\
Grok-4     & 40.5 (27.6, 53.4) & 30.1 (21.3, 38.8) & 38.8 (31.0, 46.7) & 40.8 (30.5, 51.1) & 32.1 (19.6, 44.5) & 24.7 (17.7, 31.7) \\
Gemini     & 52.4 (40.2, 64.6) & 30.6 (21.2, 40.0) & 34.1 (26.8, 41.4) & 33.0 (22.8, 43.2) & 26.3 (15.5, 37.1) & 10.3 (5.5, 15.0) \\
\bottomrule
\end{tabular}
\end{adjustbox}
\end{table}

\begin{figure}[t]
\centering
\includegraphics[width=\textwidth]{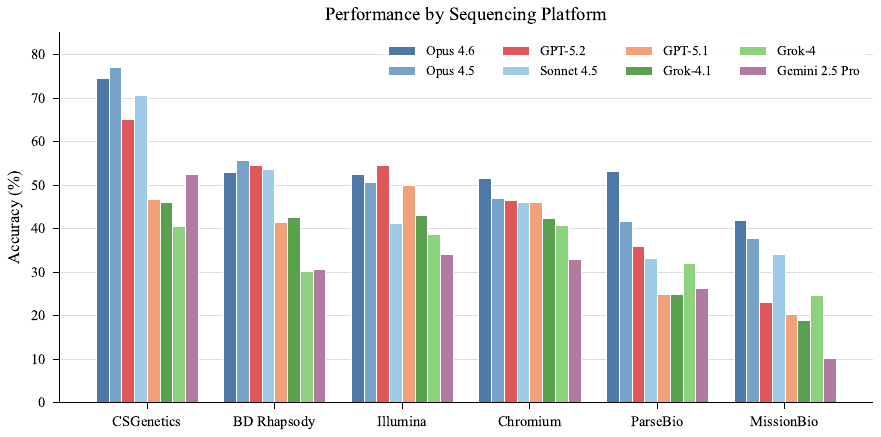}
\caption{Accuracy (\%) by sequencing platform. Platforms ordered by decreasing cross-model mean accuracy. Error bars show 95\% confidence intervals.}
\label{fig:platform}
\end{figure}

Platform choice affects accuracy as much as model choice (Table~\ref{tab:byplatform}, Figure~\ref{fig:platform}). Cross-model mean accuracy ranges from 59.1\% on CSGenetics to 26.4\% on MissionBio---a 32.7 pp gap that exceeds the 23.6 pp spread between best and worst models. CSGenetics is easiest for six of eight models; MissionBio is hardest for all eight.

MissionBio inverts rankings. Grok-4 (sixth overall) beats GPT-5.2 (third
overall) on MissionBio (24.7\% vs 23.0\%), and Sonnet 4.5 surpasses GPT-5.2 by
11 pp. The Anthropic models hold up on MissionBio while most competitors
collapse.

Every model shows large platform swings. Gemini drops 42 pp between CSGenetics
(52.4\%) and MissionBio (10.3\%). Even Opus 4.5, the most consistent model,
loses 39 pp between its best and worst platforms. These effects likely reflect
uneven training data: e.g., MissionBio appears less frequently in public
documentation than Chromium pipelines.

\subsection{Comparison to Spatial Transcriptomics}

\begin{table}[t]
\caption{Comparison of scBench (scRNA-seq) and SpatialBench (spatial transcriptomics) under the mini-SWE-agent harness.}
\label{tab:comparison}
\centering
\small
\begin{tabular}{lcc}
\toprule
 & scBench & SpatialBench \\
\midrule
Number of evaluations     & 394       & 146 \\
Number of platforms       & 6         & 5 \\
Number of task categories & 7         & 7 \\
Top model accuracy        & 52.8\%    & 38.4\% \\
Bottom model accuracy     & 29.2\%    & 20.1\% \\
Top--bottom spread        & 23.6 pp   & 18.3 pp \\
Easiest task (best model) & Norm.\ 84\% & Norm.\ 76\% \\
Hardest task (best model) & DE 41\%   & QC 22\% \\
\bottomrule
\end{tabular}
\end{table}

\begin{figure}[t]
\centering
\includegraphics[width=\textwidth]{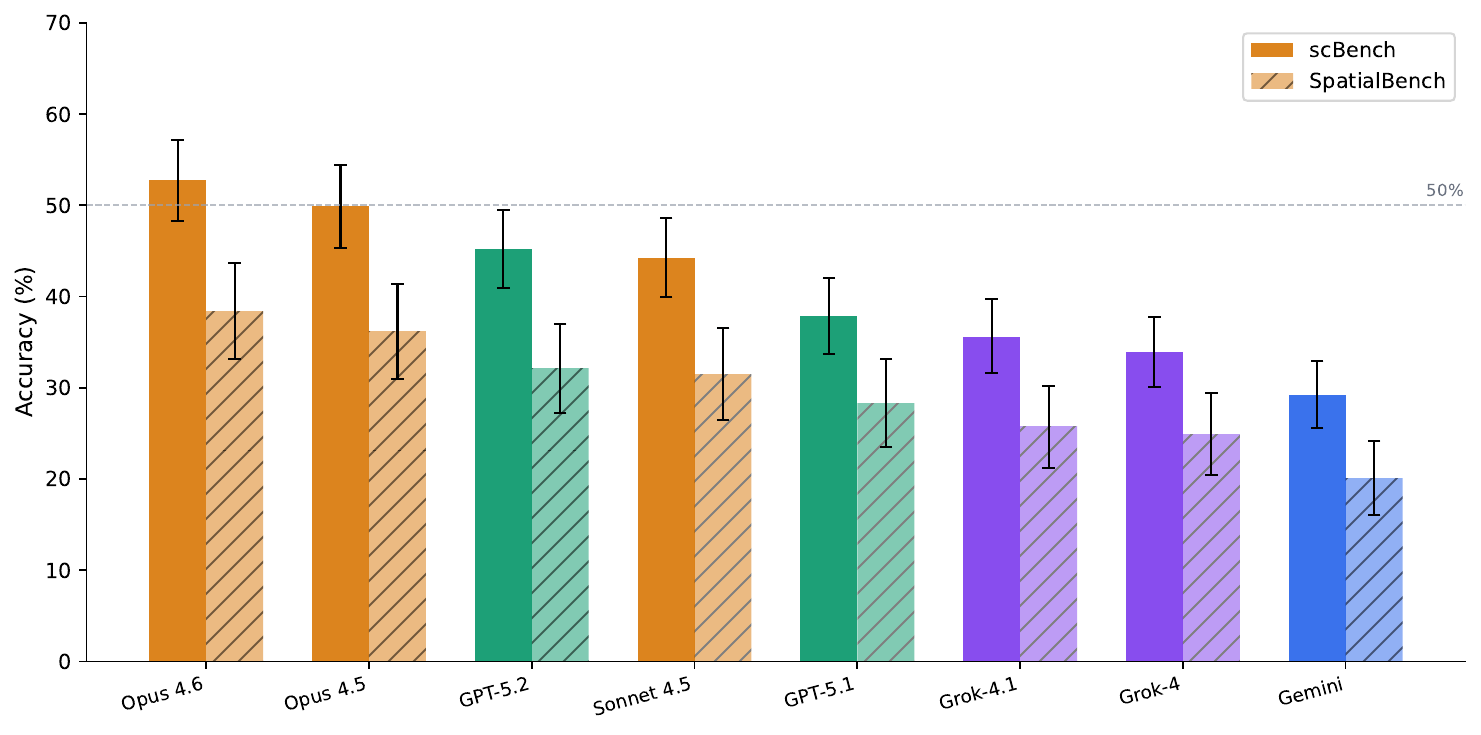}
\caption{Model accuracy on scBench (solid bars) versus SpatialBench (hatched bars). scRNA-seq yields consistently higher accuracy across all models, but rankings are preserved: Claude Opus leads both benchmarks, Gemini ranks last. Error bars show 95\% CIs.}
\label{fig:comparison}
\end{figure}

scBench and SpatialBench~\citep{workman2025spatialbench} together cover the two
dominant transcriptional assays (Table~\ref{tab:comparison}). The top model
reaches 52.8\% on scBench versus 38.4\% on SpatialBench---scRNA-seq is more
tractable. This gap holds across the leaderboard: the bottom model scores
29.2\% on scBench versus 20.1\% on SpatialBench. Model rankings are preserved
at the extremes: Claude Opus leads both benchmarks and Gemini ranks last in
both.

The benchmarks share some structure. Normalization is easiest in both (84\% vs
76\% best-model accuracy). Both show strong platform effects, with 30--40 pp
swings depending on sequencing technology. The accuracy gap likely also
reflects training data: scRNA-seq has far more public datasets than spatial
transcriptomics, and tools like Scanpy dominate the ecosystem with extensive
documentation. This is perhaps most obvious in the difference in performance on
the QC task category, where knowledge of thresholds and other procedurally
simple operations explains variability.

\section{Discussion}

Agents for scRNA-seq occupy the same capability regime that SpatialBench
exposed for spatial transcriptomics. While they demonstrate some capability,
they are unable to faithfully extract biological insight from messy, real-world
datasets. Across 394 verifiable problems with deterministic grading, the best
model reaches 52.8\% accuracy, leaving substantial room for progress. The
23.6-point spread across models shows that scBench discriminates capability,
with significant task- and platform-dependent behavioral shifts. In practice,
these results suggest that today's agents can accelerate routine analysis but
cannot yet be trusted to autonomously answer scientific questions without
stringent verification of intermediate results and human oversight.

As with SpatialBench, the path forward appears to be a long tail of tractable
engineering. Tasks that demand contextual, often tacit judgment remain the
least reliable. Normalization and QC are approaching reliability, while cell
typing and differential expression require contextual decision-making and
scientific reasoning currently outside the capabilities of frontier models.
General-purpose coding skill is not sufficient; models need exposure to
representative scRNA-seq workflows across diverse tissue and disease contexts,
in addition to thorough understanding of technology-specific analysis
techniques.

Platform-dependent performance swings often exceed task-dependent ones,
suggesting that reliable agents will require platform-aware context,
assay-specific tooling, and self-calibration heuristics rather than
one-size-fits-all reasoning. The MissionBio collapse and the Illumina-specific
strength of certain models reflect gaps in training data and the fragility of
memorized techniques when confronted with unfamiliar problems.

scBench shares SpatialBench's limitations: deterministic graders enable
verifiable evaluation but necessarily discretize scientific judgment into
automatically checkable chunks, and each evaluation snapshots a single
workflow step rather than capturing long-horizon iteration where errors
compound and thresholds are revisited. We hope scBench serves both as a
measurement tool and a diagnostic lens, an evolving specification of scRNA-seq
competence that supports test-driven development of agent systems whose
behavior can improve through both model training and harness engineering.

\section{Methods}

\subsection{Problem Construction}

scBench is constructed from real scRNA-seq analysis workflows across six sequencing platforms: Chromium, BD Rhapsody, CSGenetics, Illumina, MissionBio, and ParseBio. Following SpatialBench, we identify analysis steps that satisfy three criteria: (1) the task arises in routine practice---a step that a working bioinformatician would perform as part of a standard pipeline (QC, normalization, HVG selection, clustering, annotation, or differential expression); (2) the answer requires empirical data interaction---it depends on the provided dataset and cannot be produced from textbook knowledge or memorized gene lists alone; and (3) the result is a verifiable quantitative artifact---a structured JSON output that can be graded deterministically by one of five grader families (Section~\ref{sec:graders}).

Each candidate problem follows a five-stage construction pipeline. We first reproduce the target analysis step on the provided data using the published workflow. We then define the output artifact as an exact JSON schema with named fields and value types. Next, we select the grader family matching the output shape (e.g., \texttt{NumericTolerance} for cell counts, \texttt{DistributionComparison} for cell type proportions). We calibrate tolerances by running the analysis with multiple valid methods and parameter choices to establish the range of acceptable answers. Finally, we harden against shortcuts by removing precomputed embeddings, cached labels, and any fields that would allow the agent to read the answer without performing the intended computation.

Ground truth values are derived by re-running published pipelines from raw counts using author-specified parameters where available, then verified against domain understanding of expected biological results. When papers do not uniquely specify parameters (e.g., QC threshold not reported), we use standard defaults and widen tolerances to accept the resulting variation. Each problem is assigned an evaluation type---scientific, procedural, or observational (Section~\ref{sec:evaltypes})---governing how aggressively tolerances must accommodate methodological variation. As a final quality-control step, we attempt to solve each problem by (a) reading \texttt{.obs} and \texttt{.uns} fields directly without computation, (b) answering from prior biological knowledge, and (c) running with alternative valid methods to verify tolerance coverage. Problems failing any of these checks are revised or removed.

\subsection{Anatomy of a Problem}
\label{sec:anatomy}

Each evaluation is a JSON specification with four agent-visible components and one internal component. The \emph{data node} points to one or more AnnData \texttt{.h5ad} files~\citep{wolf2018scanpy} containing the expression matrix, cell metadata (\texttt{.obs}), and gene annotations (\texttt{.var}); at runtime the harness downloads these files into an isolated workspace. The \emph{task prompt} describes the analysis goal in natural language and specifies the exact JSON output format, including field names and value types. The \emph{deterministic grader} defines the grader family and its configuration---ground truth values, tolerance parameters, and pass thresholds---that map the agent's structured answer to pass/fail (Section~\ref{sec:graders}). \emph{Metadata} tags each problem by task category, evaluation type, sequencing platform, and computational complexity. A fifth component, \emph{notes}, documents the solution approach, tolerance rationale, and known edge cases; notes are excluded from the agent's context at the harness level and never appear at runtime.

Eval definitions are validated by a deterministic \emph{linter} before entering the benchmark. The linter performs static schema validation, checking that required fields are present, grader configurations are well-formed (e.g., tolerance types are valid, thresholds are in range), and that the answer fields specified in the task prompt match what the grader expects. Evals that fail linting are blocked; evals with ambiguous tolerances or shortcut-prone structure are revised or removed during manual review.

\subsection{Evaluation Types and Durability}
\label{sec:evaltypes}

Every evaluation is classified into one of three types that govern how aggressively tolerances must accommodate methodological variation.

\paragraph{Scientific.} The prompt specifies a biological goal but leaves both the method and its parameters to the agent (e.g., ``filter low-quality cells''). Because multiple QC thresholds, HVG selection methods, or clustering resolutions could defensibly be applied to the same data, tolerances must be wide enough to accept all reasonable choices. Tight tolerances are used only when clean data causes valid methods to converge.

\paragraph{Procedural.} The prompt names a specific method and leaves only parameter choices to the agent (e.g., ``normalize using scran pooling''~\citep{lun2016pooling}). Tolerances can be tighter than for scientific evaluations because the method is constrained.

\paragraph{Observational.} The prompt asks the agent to interpret or report a property of the data (e.g., ``which cell populations separate along PC1?''). Durability requirements are relaxed, and grading focuses on verifiability and anti-shortcut structure.

\medskip
\noindent The distribution of evaluation types affects aggregate interpretation: a benchmark weighted toward procedural evaluations would yield higher scores because the method is specified, while a scientific-heavy benchmark tests judgment under ambiguity. Scientific evaluations carry wider tolerances on average than procedural evaluations, reflecting the greater methodological freedom.

\subsection{Design Principles}
\label{sec:principles}

Following SpatialBench, we apply three design criteria to every evaluation. The overarching rule is \emph{specify what, not how}: tasks define the scientific goal and the exact output format, but do not prescribe a step-by-step method or parameters (with the exception of procedural evaluations, which name the method). The linter enforces structural compliance; manual review validates each criterion.

\paragraph{Verifiability.} Each task specifies an exact JSON output format with named fields and value types, and is paired with a deterministic grader whose output shape matches the task (e.g., \texttt{NumericTolerance} for cell counts, \texttt{DistributionComparison} for cell type proportions). Success is automatically checkable with no subjective interpretation. Tasks that rely on subjective language (``interesting'', ``meaningful'') without an operational definition are rejected. Importantly, omitting thresholds or algorithm names is acceptable and often desirable---it preserves anti-shortcut structure by forcing the agent to make data-driven choices.

\paragraph{Scientific durability.} The intended answer must be stable across reasonable methodological choices, or tolerances must be wide enough to accept the resulting variation. Durability requirements scale with evaluation type (Section~\ref{sec:evaltypes}): scientific evaluations demand the widest tolerances, procedural evaluations can be tighter, and observational evaluations are the most relaxed. We specifically avoid two failure modes common in scRNA-seq benchmarking: random seed sensitivity (Leiden clustering~\citep{traag2019leiden} is stochastic, so we do not test exact cluster counts) and library version artifacts (UMAP~\citep{mcinnes2018umap} coordinates are arbitrary across versions, so we test biological interpretation rather than coordinates). Imprecision that a domain expert would resolve unambiguously (e.g., ``filter low-quality cells'' without specifying which metric) is not considered a durability failure; only ambiguity where reasonable interpretations yield materially different answers is flagged.

\paragraph{Anti-shortcut.} The agent must load and analyze the provided dataset; prior knowledge alone is insufficient. During problem construction, we remove precomputed embeddings (\texttt{adata.obsm["X\_pca"]}, \texttt{adata.obsm["X\_umap"]}) when the evaluation tests computation, strip cached labels and summary statistics that would allow the agent to read the answer directly, and ensure that multiple-choice distractors are biologically plausible to prevent label leakage. That an answer is ``just a number'' does not make it guessable---dataset-specific quantities are not shortcuttable once precomputed fields are removed.

\subsection{Graders}
\label{sec:graders}

Each evaluation is paired with a deterministic grader that maps the agent's structured JSON answer to pass/fail. We use five grader families; formal specifications are in Appendix~C.

\paragraph{NumericTolerance.} Validates numeric values such as cell counts, expression levels, and QC metrics. Supports four tolerance modes---absolute ($|x - x^*| \leq \epsilon$), relative ($|x - x^*|/|x^*| \leq \epsilon$), minimum ($x \geq x_{\min}$), and maximum ($x \leq x_{\max}$)---as well as asymmetric bounds. Multiple fields are checked independently; all must pass. String values are coerced to floats; coercion failure counts as a field failure.

\paragraph{MultipleChoice.} Validates discrete answers against one or more correct options. The agent's response is trimmed and uppercased before comparison, making matching case-insensitive.

\paragraph{MarkerGenePrecisionRecall.} Validates gene lists against canonical marker sets using recall@$K$ (fraction of canonical markers recovered) and precision@$K$ (fraction of returned genes that are canonical). Gene names are lowercased before comparison. Recall thresholds are set per evaluation (typically $\geq 0.50$); precision thresholds default to $\geq 0.60$ but are set to zero when the evaluation tests recall without penalizing novel DE genes. A per-cell-type mode supports multi-population differential expression by requiring a minimum recall for each cell type.

\paragraph{LabelSetJaccard.} Validates unordered set predictions (e.g., predicted cell type labels) via the Jaccard index $J(A,B) = |A \cap B|\,/\,|A \cup B|$, with a default pass threshold of $0.90$. Both missing and extra labels penalize the score equally. Labels are compared as-is without case normalization.

\paragraph{DistributionComparison.} Validates multi-category proportions such as cell type distributions. Each ground-truth category is checked independently against an absolute tolerance (e.g., $\pm 5$ percentage points); all categories must pass. Categories missing from the agent's output fail automatically, while extra categories are ignored. Category names are lowercased before comparison. The all-must-pass rule ensures that agents cannot ignore rare cell types; tolerances are set wide enough to absorb reasonable per-category variation.

\subsection{Agent Harness}

We evaluate all models under mini-SWE-agent~\citep{yang2024sweagent}, an open-source harness that implements a simple action loop: the LLM generates a free-form response, the harness extracts the first fenced code block (delimited by markdown triple-backtick syntax), executes it in a local bash shell, and returns stdout/stderr to the model as the next observation. Each evaluation is capped at 100 action steps (LLM turn $\to$ code extraction $\to$ execution $\to$ observation); if the agent exhausts the step budget without writing \texttt{eval\_answer.json}, the evaluation scores zero.

The runtime environment provides scanpy, anndata, numpy, pandas, scipy, and matplotlib; all models share the same package versions. Network access is enabled, allowing agents to install additional packages if needed. Each evaluation runs in an isolated workspace: data files are symlinked from a local cache and the agent has read/write access only within that workspace. No GUI or interactive tools (Jupyter, plot display) are available.

Two timeout layers bound execution. An operation timeout of 300 seconds caps any individual bash command; an evaluation timeout of 600 seconds (configurable per evaluation) caps total wall-clock time via \texttt{SIGALRM}. On timeout or runtime crash, the harness grades whatever is in \texttt{eval\_answer.json} at that point; if no answer file exists, the evaluation scores zero. There is no retry logic---each replicate is a single attempt. The harness records a complete trajectory for every run (conversation history, tool calls, and outputs), enabling post-hoc analysis of agent behavior.

\subsection{Statistical Design}

We follow the same two-stage aggregation used in SpatialBench. Each model--evaluation pair is run $K{=}3$ times. Replicates share the same prompt, data, and harness; the only source of variation is the model's sampling nondeterminism (no explicit seed or temperature control). Each run receives a binary outcome from the grader, $s_{i,r} \in \{0, 1\}$.

In the first stage we compute the per-evaluation mean $\bar{s}_i = \frac{1}{K}\sum_{r} s_{i,r}$, yielding a value in $\{0, \tfrac{1}{3}, \tfrac{2}{3}, 1\}$. In the second stage we treat the $\{\bar{s}_i\}_{i=1}^{n}$ as independent observations and compute the aggregate accuracy $\hat{\mu} = \frac{1}{n}\sum_i \bar{s}_i$ with 95\% confidence intervals via the $t$-distribution on $n{-}1$ degrees of freedom. All 394 evaluations are equally weighted; there is no upweighting by task category or platform. Per-evaluation means are approximately independent because evaluations use different datasets and different prompts; the shared model is the only common factor and is constant within a model's column. For stratified breakdowns (by task or platform), we apply the same procedure to the relevant subset, recomputing $n$, $\hat{\mu}$, and the corresponding $t$ critical value.

\section{Data and Code Availability}

The benchmark framework, graders, linter, and agent harness are available at \url{https://github.com/latchbio/scbench}. Thirty canonical evaluations across five platforms (Chromium, CSGenetics, Illumina, MissionBio, ParseBio) are publicly released to demonstrate the benchmark format, along with full agent trajectories for all canonical evaluations. The full 394-evaluation suite is withheld to prevent training contamination. Aggregate results (per-model, per-task, per-platform breakdowns) are included in the repository. The evaluation framework supports custom agents via a pluggable \texttt{agent\_function} interface, enabling direct comparison of new models against the published results.

\section*{Author Contributions}

H.M., Z.Y., and H.L.\ wrote evaluations and collected data. K.W.\ and A.A.\ built the evaluation tooling. K.W.\ wrote the manuscript.

\vspace{1em}
\noindent\textit{Dedicated to AM.}

\appendix

\section*{A. Benchmark Inventory}
\label{app:inventory}
\addcontentsline{toc}{section}{A. Benchmark Inventory}

scBench comprises 394 evaluations drawn from published analyses on six sequencing platforms. Each platform uses a distinct library preparation and capture technology, ensuring that the benchmark tests generalization across the scRNA-seq ecosystem rather than proficiency on a single data format.

\begin{itemize}
\item \textbf{BD Rhapsody}~\citep{shum2019rhapsody}: microwell-based capture with targeted or whole-transcriptome panels. 61 evaluations.
\item \textbf{Chromium}~\citep{zheng2017chromium} (10x Genomics): droplet-based capture. The most widely used scRNA-seq platform and the best represented in public datasets and documentation. 60 evaluations.
\item \textbf{CSGenetics}~\citep{csgenetics_simplecell}: droplet-based capture with a proprietary barcoding chemistry. 42 evaluations.
\item \textbf{Illumina}~\citep{picelli2014smartseq2}: plate-based single-nucleus RNA-seq (DRG tissue). 85 evaluations.
\item \textbf{MissionBio}~\citep{ruff2022tapestri} (Tapestri): targeted panel sequencing of DNA, RNA, and surface protein. Non-standard data structures and less common analysis tooling make this the hardest platform in the benchmark. 81 evaluations.
\item \textbf{ParseBio}~\citep{rosenberg2018splitseq}: split-pool combinatorial barcoding (no microfluidics). 65 evaluations.
\end{itemize}

\noindent Tissue types span PBMCs, tumor microenvironments (4T1 mammary carcinoma, CDX models), dorsal root ganglia (DRG), and hematopoietic samples. Table~\ref{tab:inventory} shows the distribution of evaluations across platforms and task categories.

\begin{table}[h]
\centering
\caption{Summary of scBench evaluations.}
\label{tab:inventory_summary}
\small
\begin{tabular}{lclc}
\toprule
\textbf{By Platform} & \textbf{Evals} & \textbf{By Task Category} & \textbf{Evals} \\
\midrule
BD Rhapsody & 61 & QC & 36 \\
Chromium    & 60 & Normalization & 44 \\
CSGenetics  & 42 & Dimensionality Reduction & 69 \\
Illumina    & 85 & Clustering & 49 \\
MissionBio  & 81 & Cell Typing & 118 \\
ParseBio    & 65 & Differential Expression & 71 \\
            &    & Trajectory Analysis & 7 \\
\midrule
\textbf{Total} & \textbf{394} & \textbf{Total} & \textbf{394} \\
\bottomrule
\end{tabular}
\end{table}

\subsection*{Grader Distribution}

Evaluations use five grader families to assess agent outputs:
\begin{itemize}
\item \textbf{NumericTolerance}: QC metrics, cell counts, expression values, fold changes (most common)
\item \textbf{MultipleChoice}: Biological interpretation, pattern identification
\item \textbf{MarkerGenePrecisionRecall}: Marker discovery, differential expression gene lists
\item \textbf{LabelSetJaccard}: Cell type prediction, cluster composition
\item \textbf{DistributionComparison}: Cell type proportions, population distributions
\end{itemize}

\subsection*{Tissue Coverage}

The benchmark covers four primary tissue/sample types:
\begin{itemize}
\item \textbf{PBMC} (BD Rhapsody, CSGenetics, ParseBio): 168 evaluations --- T cell subtypes, monocyte populations, rare cell detection
\item \textbf{Tumor microenvironment} (Chromium): 60 evaluations --- 4T1 mammary carcinoma, CDX small-cell lung cancer, CAF subtypes
\item \textbf{Dorsal root ganglia} (Illumina): 85 evaluations --- neuron subclasses, satellite glial cells, age-related changes
\item \textbf{Hematopoietic} (MissionBio): 81 evaluations --- CCUS samples, clonal hierarchy, mutation burden
\end{itemize}

\subsection*{Complete Evaluation Inventory}

Table~\ref{tab:full-inventory} provides the complete list of all 394 evaluations organized by platform.

\begin{small}
\begin{longtable}{p{4cm}p{1.8cm}p{1.5cm}l}
\caption{Complete inventory of scBench evaluations.}
\label{tab:full-inventory} \\
\toprule
Description & Platform & Task & Grader \\
\midrule
\endfirsthead
\multicolumn{4}{c}{\tablename\ \thetable{} -- continued} \\
\toprule
Description & Platform & Task & Grader \\
\midrule
\endhead
\midrule
\multicolumn{4}{r}{Continued on next page} \\
\endfoot
\bottomrule
\endlastfoot
Naive T cell marker comparison & BD Rhapsody & Cell Typ. & MCQ \\
Treg marker gene recall & BD Rhapsody & Cell Typ. & P@K \\
CD8 TEM vs naive classification & BD Rhapsody & Cell Typ. & MCQ \\
Effective subtype count & BD Rhapsody & Cell Typ. & Numeric \\
Baseline iNKT fraction & BD Rhapsody & Cell Typ. & Numeric \\
CD8 TEM trend contrast & BD Rhapsody & Cell Typ. & MCQ \\
Classical monocyte pattern & BD Rhapsody & Cell Typ. & MCQ \\
CD14 score separation & BD Rhapsody & Cell Typ. & Numeric \\
Proliferative lymphocyte rarity & BD Rhapsody & Cell Typ. & Numeric \\
Subtype stability under reclustering & BD Rhapsody & Cell Typ. & Numeric \\
Patient composition divergence & BD Rhapsody & Cell Typ. & Numeric \\
21-subtype distribution (v1) & BD Rhapsody & Cell Typ. & Dist \\
21-subtype distribution (v2) & BD Rhapsody & Cell Typ. & Dist \\
Marker program coverage & BD Rhapsody & Clust. & P@K \\
Cytotoxic program cluster & BD Rhapsody & Clust. & MCQ \\
Cluster count & BD Rhapsody & Clust. & Numeric \\
Program separation overlap & BD Rhapsody & Clust. & Numeric \\
Subtype expression shift & BD Rhapsody & Clust. & Numeric \\
S100A vs MHC enrichment & BD Rhapsody & Clust. & Numeric \\
Louvain resolution sweep & BD Rhapsody & Clust. & Numeric \\
Day 3 stress gene fraction & BD Rhapsody & DE & Numeric \\
CD4 TEM EGR1 log fold change & BD Rhapsody & DE & Numeric \\
CD4 TEM RGS1 log fold change & BD Rhapsody & DE & Numeric \\
CD14 monocyte TNFa log fold change & BD Rhapsody & DE & Numeric \\
IFITM3 temporal pattern & BD Rhapsody & DE & MCQ \\
IL1B temporal pattern & BD Rhapsody & DE & MCQ \\
FCER1G day 3 expression & BD Rhapsody & DE & MCQ \\
Adhesion gene return to baseline & BD Rhapsody & DE & MCQ \\
DE temporal pattern (09) & BD Rhapsody & DE & MCQ \\
DE temporal pattern (10) & BD Rhapsody & DE & MCQ \\
Dimensionality reduction (14 evals) & BD Rhapsody & Dim.Red. & Mixed \\
Normalization (11 evals) & BD Rhapsody & Norm. & Numeric \\
Quality control (6 evals) & BD Rhapsody & QC & Numeric \\
\midrule
CAF subcluster cell typing (5 evals) & Chromium & Cell Typ. & Mixed \\
Tumor clustering (6 evals) & Chromium & Clust. & Mixed \\
Contractile CAF marker recovery & Chromium & DE & P@K \\
Differential expression (10 evals) & Chromium & DE & Mixed \\
Dimensionality reduction (15 evals) & Chromium & Dim.Red. & Mixed \\
Normalization (11 evals) & Chromium & Norm. & Numeric \\
Quality control (10 evals) & Chromium & QC & Numeric \\
\midrule
PBMC cell type proportions & CSGenetics & Cell Typ. & Dist \\
T cell marker recovery & CSGenetics & Cell Typ. & P@K \\
T cell activation/exhaustion state & CSGenetics & Cell Typ. & MCQ \\
Cell typing (17 additional evals) & CSGenetics & Cell Typ. & Mixed \\
Clustering (5 evals) & CSGenetics & Clust. & Mixed \\
Differential expression (1 eval) & CSGenetics & DE & Numeric \\
Dimensionality reduction (7 evals) & CSGenetics & Dim.Red. & Mixed \\
Normalization (5 evals) & CSGenetics & Norm. & Numeric \\
Quality control (4 evals) & CSGenetics & QC & Numeric \\
\midrule
Neuron subclass assignment & Illumina & Cell Typ. & Jaccard \\
Brain signature in DRG (adversarial) & Illumina & Cell Typ. & MCQ \\
Cell typing (31 additional evals) & Illumina & Cell Typ. & Mixed \\
Leiden cluster count & Illumina & Clust. & Numeric \\
Clustering (11 additional evals) & Illumina & Clust. & Mixed \\
Differential expression (15 evals) & Illumina & DE & Mixed \\
Dimensionality reduction (10 evals) & Illumina & Dim.Red. & Mixed \\
Normalization (7 evals) & Illumina & Norm. & Numeric \\
Quality control (8 evals) & Illumina & QC & Numeric \\
\midrule
Cell type label set & MissionBio & Cell Typ. & Jaccard \\
Other cell fraction & MissionBio & Cell Typ. & Numeric \\
NK marker recovery (top 5) & MissionBio & Cell Typ. & P@K \\
Cell typing (19 additional evals) & MissionBio & Cell Typ. & Mixed \\
CCUS clonal typing (10 evals) & MissionBio & Cell Typ. & MCQ \\
Louvain cluster count & MissionBio & Clust. & Numeric \\
Clustering (11 additional evals) & MissionBio & Clust. & Mixed \\
Differential expression (19 evals) & MissionBio & DE & Mixed \\
Dimensionality reduction (5 evals) & MissionBio & Dim.Red. & Mixed \\
Normalization (3 evals) & MissionBio & Norm. & Mixed \\
Quality control (8 evals) & MissionBio & QC & Numeric \\
\midrule
cDC2 annotation confusion & ParseBio & Cell Typ. & MCQ \\
Cell typing (12 additional evals) & ParseBio & Cell Typ. & Mixed \\
Clustering (5 evals) & ParseBio & Clust. & Mixed \\
IL-4 monocyte response & ParseBio & DE & Numeric \\
Differential expression (21 evals) & ParseBio & DE & Mixed \\
Dimensionality reduction (18 evals) & ParseBio & Dim.Red. & Mixed \\
Normalization (7 evals) & ParseBio & Norm. & Numeric \\
\end{longtable}
\end{small}

\noindent\textit{Grader abbreviations: MCQ = MultipleChoice, P@K = MarkerGenePrecisionRecall, Numeric = NumericTolerance, Jaccard = LabelSetJaccard, Dist = DistributionComparison. Full evaluation specifications are available in the benchmark repository.}

\section*{B. Canonical Examples}
\label{app:examples}
\addcontentsline{toc}{section}{B. Canonical Examples}

We provide 2--3 representative evaluations from each platform to illustrate the benchmark format and grader diversity, with emphasis on downstream analysis tasks (cell typing, clustering, differential expression). For each we list the task category, grader type, and tolerance rationale.

\subsection*{BD Rhapsody}

\paragraph{Cell Typing (MarkerGenePrecisionRecall).}
\texttt{bd\_rhapsody\_celltyping\_02\_treg\_...}. The agent identifies marker genes for regulatory T cells from PBMC data. Canonical markers: \emph{FOXP3}, \emph{IL2RA}, \emph{CTLA4}, \emph{DUSP4}, \emph{RGS1} (5 total). Pass: recall@10 $\geq 0.60$, precision $\geq 0$.

\paragraph{Clustering (NumericTolerance).}
\texttt{bd\_rhapsody\_clustering\_03\_count}. The agent clusters PBMC cells and reports the number of clusters. Ground truth: 12 clusters; tolerance $\pm 2$ (absolute). The tolerance accommodates variation across resolution parameters and clustering algorithms.

\subsection*{Chromium}

\paragraph{Cell Typing (LabelSetJaccard).}
\texttt{chromium\_celltyping\_03\_caf\_subcluster\_...}. The agent subclusters cancer-associated fibroblasts (CAFs) from a 4T1 tumor dataset and identifies marker programs for each subcluster. Ground truth: 12 canonical markers including \emph{Acta2}, \emph{Col1a1}, \emph{Mki67}, \emph{Ly6c1}. Pass: Jaccard $\geq 0.60$.

\paragraph{Differential Expression (MarkerGenePrecisionRecall).}
\texttt{chromium\_de\_01\_...}. The agent subclusters CAFs, identifies the contractile subcluster, and reports top 50 DE genes. Canonical markers include \emph{Tpm1}, \emph{Myl9}, \emph{Tagln} (9 total). Pass: recall $\geq 0.67$.

\paragraph{Clustering (NumericTolerance).}
\texttt{chromium\_cdx\_sclc\_heterogeneity\_...}. The agent computes intra-cluster heterogeneity before and after cisplatin treatment in a CDX small-cell lung cancer model, reporting the fold change. Pass: fold change $\geq 1.0$ (minimum threshold).

\subsection*{CSGenetics}

\paragraph{Cell Typing (DistributionComparison).}
\texttt{pbmc\_cell\_type\_annotation\_v1}. The agent annotates PBMC cells into five compartments (T cells, B cells, NK cells, Monocytes, Dendritic cells) and reports percentages. Ground truth: 59.0\%/20.3\%/5.0\%/14.8\%/0.7\%; tolerance $\pm 5$ pp per category.

\paragraph{Cell Typing (MarkerGenePrecisionRecall).}
\texttt{pbmc\_t\_cells\_marker\_recovery}. The agent identifies the top 20 marker genes for T cells. Canonical markers include \emph{CD3D}, \emph{CD3E}, \emph{IL7R}, \emph{TRAC}, \emph{TRBC2} (10 total). Pass: recall $\geq 0.50$, precision $\geq 0$.

\paragraph{Cell Typing (MultipleChoice).}
\texttt{pbmc\_tcell\_dual\_activation\_exhaustion}. The agent interprets T cell states to determine which subpopulation shows both activation and exhaustion signatures. Correct answer: H. Requires understanding of T cell biology beyond marker lookup.

\subsection*{Illumina}

\paragraph{Cell Typing (LabelSetJaccard).}
\texttt{snrna\_anno\_03\_assign\_neuron\_subclasses\_...}. The agent assigns neuron subclasses (NF, NP, PEP, TH) in DRG snRNA-seq data based on marker expression. Ground truth: 4 neuron subclasses. Pass: Jaccard $\geq 0.80$.

\paragraph{Clustering (NumericTolerance).}
\texttt{snrna\_ic\_11\_leiden\_cluster\_and\_report\_n\_clusters}. The agent applies Leiden clustering to DRG snRNA-seq data and reports the cluster count. Ground truth: 16 clusters; tolerance $\pm 3$ (absolute).

\paragraph{Cell Typing (MultipleChoice).}
\texttt{snrna\_anno\_adv\_forced\_03\_...\_brain\_region\_bait}. An adversarial forced-choice evaluation: the agent is given DRG (peripheral nervous system) data but asked which brain-region-specific neuronal signature scores highest. The correct answer (F, Microglia\_homeostatic) appears because of shared macrophage-like markers, not because microglia are present in DRG. Tests whether the agent can detect that brain signatures are biologically implausible in this tissue context.

\subsection*{MissionBio}

\paragraph{Cell Typing (MarkerGenePrecisionRecall).}
\texttt{annotation\_03\_nk\_marker\_recovery\_top5}. Using the Tapestri protein panel, the agent identifies the top 5 markers for NK cells. Canonical markers: \emph{CD16}, \emph{CD56}. Pass: recall $\geq 0.50$, precision $\geq 0.40$.

\paragraph{Cell Typing (MultipleChoice).}
\texttt{ccus\_ct\_09\_highest\_mutation\_burden\_...}. Using the Tapestri multi-omic panel, the agent identifies which cell population carries the highest per-cell mutation burden. Correct answer: A. Requires integrating DNA variant calls with cell labels.

\subsection*{ParseBio}

\paragraph{Cell Typing (MultipleChoice).}
\texttt{pbmc\_cdc2\_annotation\_confusion}. The agent resolves an annotation ambiguity in a split-pool PBMC dataset by interpreting marker overlap between cDC2 and other myeloid populations. Correct answer: A.

\paragraph{Differential Expression (NumericTolerance).}
\texttt{parsebio\_il4\_monocyte\_response}. The agent computes the log2 fold change of a target gene in monocytes under IL-4 stimulation. Ground truth: $-1.25$; pass if log2FC $\leq -1.1$ (maximum threshold, directional).

\section*{C. Grader Specification}
\label{app:graders}
\addcontentsline{toc}{section}{C. Grader Specification}

The graders, linter, and harness are implemented in the open-source \texttt{latchbio/latch-eval-tools} repository.\footnote{\url{https://github.com/latchbio/latch-eval-tools}} Below we give formal specifications for each grader family.

\subsection*{NumericTolerance}

\textbf{Input:} JSON object with one or more numeric fields. String values are coerced via \texttt{float()}; coercion failure counts as a field failure.

\textbf{Tolerance modes} (configured per field):
\begin{itemize}
\item Absolute: pass if $|x - x^*| \leq \epsilon$
\item Relative: pass if $|x - x^*|/|x^*| \leq \epsilon$
\item Minimum: pass if $x \geq x_{\min}$
\item Maximum: pass if $x \leq x_{\max}$
\item Asymmetric: pass if $x^* - \epsilon_{\text{lower}} \leq x \leq x^* + \epsilon_{\text{upper}}$
\end{itemize}
Multiple fields are checked independently; all must pass. Missing required fields fail. Extra keys are ignored.

\subsection*{MultipleChoice}

\textbf{Input:} \verb|{"answer": "A"}|.

\textbf{Normalization:} agent answer is trimmed and uppercased (\texttt{.strip().upper()}).

\textbf{Pass criterion:} agent answer is a member of the configured \texttt{correct\_answers} list.

\subsection*{MarkerGenePrecisionRecall}

\textbf{Input:} a gene list (flat mode) or a dictionary mapping cell types to gene lists (per-cell-type mode). Gene names are lowercased before comparison.

\textbf{Flat mode.} Let $P$ be the agent's gene set and $G$ the canonical marker set.
\[
\text{precision@}K = \frac{|P \cap G|}{|P|}, \qquad \text{recall@}K = \frac{|P \cap G|}{|G|}
\]
Pass if precision $\geq \tau_p$ \textbf{and} recall $\geq \tau_r$. Defaults: $\tau_r = 0.50$, $\tau_p = 0.60$ (overridable; $\tau_p = 0$ disables precision penalty).

\textbf{Per-cell-type mode.} Recall is computed per cell type; a cell type passes if recall $\geq$ \texttt{min\_recall\_per\_celltype}. The evaluation passes if the count of passing cell types $\geq$ \texttt{min\_celltypes\_passing}.

\subsection*{LabelSetJaccard}

\textbf{Input:} a list of predicted labels. Labels are compared as-is (no case normalization).

\textbf{Pass criterion:}
$J(A, B) = |A \cap B|\,/\,|A \cup B| \geq \tau$ (default $\tau = 0.90$).
Missing and extra labels both reduce the Jaccard index.

\subsection*{DistributionComparison}

\textbf{Input:} a dictionary mapping category names to percentages. Category names are lowercased.

\textbf{Pass criterion:} for each ground-truth category $c$,
$|p_c^{\text{agent}} - p_c^{\text{gt}}| \leq \epsilon$ (default $\epsilon = 3.0$ pp).
All ground-truth categories must pass. Missing categories fail; extra categories are ignored.

\subsection*{Failure Modes}

All graders classify failures into four modes: (1) format error (missing or unparseable JSON), (2) missing field, (3) type error (coercion failure), (4) wrong value (out of tolerance). All yield score zero; the grader's reasoning field records which mode.

\bibliographystyle{plainnat}
\bibliography{references}

\end{document}